\documentclass[aps,prd,twocolumn,showkeys,floats,showpacs, nofootinbib]{revtex4-1}
\usepackage{graphicx}
\usepackage{amsmath,amssymb}
\usepackage{aas_macros}
\usepackage{natbib}
\usepackage[colorlinks=true]{hyperref}
\usepackage{hyperref}

\usepackage{bm}

\usepackage{color}

\def\af{\alpha_\textrm{EM}}

\begin{document}
\title{On dilatons with intrinsic decouplings}

\author{Olivier Minazzoli}
\email{ominazzoli@gmail.com}
\affiliation{Centre Scientifique de Monaco, 8 Quai Antoine 1er, Monaco}
\affiliation{Universit\'e C\^ote d'Azur, OCA, CNRS, Artemis, France }

\author{Aur\'elien Hees}
\email{ahees@astro.ucla.edu}
\affiliation{Department of Physics and Astronomy, University of California, Los Angeles, CA 90095, USA}

\begin{abstract}
In this paper, we show that there exists a class of dilaton models with non-trivial scalar-Ricci and scalar-matter couplings that strongly reduces observational deviations from general relativity in the dust limit. Essentially, depending on the coupling between the dilaton and the fundamental matter fields, various strengths of decoupling can appear. They range from no decoupling at all to a total decoupling state. In this latter case, the theory becomes indistinguishable from general relativity (in the dust limit), as all dilatonic effects can be re-absorbed through a simple change of unit. 

Furthermore, for particular decouplings, we show that the phenomenology used to constrain theories from universality of free fall observations is significantly different from what is commonly used. Finally, from a fundamental perspective, the class of non-dynamical decouplings proposed in this paper might play a role in the current non-observation of any deviation from general relativity (in both tests of the equivalence principle and of the parametrized post-Newtonian formalism). 
\end{abstract}

\maketitle

\section{Introduction} 
A general feature of all string and superstring theory models is the presence of a massless scalar partner to the metric called dilaton \cite{becker:2007bk,*green:1988oa,*tong:2009ar}. Perturbative calculations in the context of string theory show that the effective bosonic action should correspond to a ``general'' scalar-tensor theory \cite{callan:1986rt,damour:1994fk,*damour:1994uq}. This means that the dilaton field has gravitational strength, that it couples to the Ricci in the effective action (in the string frame) and that its kinetic term is not necessarily canonical. Besides, the dilaton is non-minimally coupled to the gauge-matter sector \cite{antoniadis:1987pl,*ferrara:1987pb,*ferrara:1987pl,*taylor:1988pb,damour:1994fk,*damour:1994uq,becker:2007bk,*green:1988oa,*tong:2009ar}. Unfortunately, the precise form of the various couplings in the effective action is not known and therefore it is impossible to make accurate predictions. However, one can derive the rich phenomenology resulting from such non-minimal couplings and constrain them with observations and experiments. 

In \cite{damour:2008pr,damour:2010zr,*damour:2010ve}, Damour and Donoghue introduced a specific  modeling for the coupling between the dilaton and matter at the microscopic level. In their model, the interaction between matter and the dilaton is parametrized by 5 dimensionless parameters that are related to fundamental parameters of the Standard Model: the fine structure constant,  the masses of the fermions (electron, quark up and quark down) and  the quantum chromodynamics (QCD) mass scale $\Lambda_3$ (see also \cite{nitti2012pr} for similar considerations). The matter Lagrangian is therefore the sum of the Standard Model Lagrangian $\mathcal L_\mathrm{SM}$ and of the interacting Lagrangian $\mathcal L_\mathrm{int}$. In natural units, this interacting Lagrangian reads
\begin{align}\label{eq:interLagrangeNL}
	\mathcal L_\mathrm{int}=& \Bigg[\frac{D_e(\varphi)}{4e^2}F_{\mu\nu}F^{\mu\nu}-\frac{D_g(\varphi)\beta_3}{2g_3}G^A_{\mu\nu}G_A^{\mu\nu}\\
	&-\sum_{i=e,u,d}(D_{m_i}(\varphi)+\gamma_{m_i}D_g(\varphi))m_i \bar \psi_i \psi_i\Bigg] \, , \nonumber
\end{align}
with $F_{\mu\nu}$ the standard electromagnetic Faraday tensor, $e$ the electric charge of the electron, $G_{\mu\nu}^A$ the gauge invariant gluon strength tensor,  $g_3$ is the QCD gauge coupling, $\beta_3$ denotes the $\beta$ function for the running of $g_3$, $m_i$ the mass of the fermions, $\gamma_{m_i}$ the anomalous dimension giving the energy running of the masses of the QCD-coupled fermions and $\psi_i$ the fermions spinor.  The functions $D_i(\varphi)$ characterize the interaction between the dilaton scalar field $\varphi$ and different matter sectors: $D_e$ characterizes the dependence of the fine structure constant to the dilaton, $D_{m_u}/D_{m_d}/D_{m_e}$ characterize the dependence of the fermions mass (quarks up and down and electron) to the dilaton and $D_g$ characterizes the dependence of the QCD mass scale $\Lambda_3$ to the dilaton (see \cite{damour:2010zr}). In this paper, following Damour and Donoghue \cite{damour:2010zr,*damour:2010ve}, we neglect strange quarks' contribution. The Lagrangian (\ref{eq:interLagrangeNL})  is a straightforward non-linear generalization of the Lagrangian from \cite{damour:2010zr,*damour:2010ve}, which is recovered when $D_i(\varphi)=d_i\varphi$ \footnote{Note that the scalar field $\varphi$ is dimensionless and is related to the scalar field $\phi$ used in \cite{damour:2010zr,*damour:2010ve} by $\varphi=\tilde\kappa\phi$ with $\tilde\kappa^2=4\pi G/c^4$.}.  We suppose here that $D_i(\varphi_0)=0$ with $\varphi_0$  the background value of the scalar field. A non-vanishing value of $D_i(\varphi_0)$ would simply lead to an unobservable rescaling of the five parameters: fine-structure constant, masses of the electron and of the up and down quarks and of the QCD mass scale $\Lambda_3$.

In~\cite{damour:2010zr,*damour:2010ve,damour:2008pr}, a simple gravitational part of the action, formed only by the Ricci scalar (i.e. the standard Einstein-Hilbert action), is used to derive observational consequences of the dilaton-matter coupling. More precisely, they introduced linear couplings between matter and the dilaton in the so-called Einstein frame. But in principle, nothing prevents to consider such linear couplings with a more general gravitational action. In particular, actual string loops expansion seems to indicate a more complicated gravitational sector in the string frame \cite{callan:1986rt,damour:1994fk,*damour:1994uq}, with a Ricci-dilaton coupling and a non-canonical kinetic term for the dilaton. 

In this paper, we show that within this general class of dilaton theories, a decoupling can arise when the dilaton-Ricci coupling is exactly related to the coupling between the dilaton and the QCD trace anomaly. This decoupling leads to a strong reduction of deviations from General Relativity (GR) for all weak field observations: violation of the Universality of Free Fall (UFF), space-time evolution of the constants of Nature, tests of the gravitational redshift and measurements of the parametrized post-Newtonian (PPN) parameters. This kind of decoupling could be part of the reasons explaining why no deviation from GR has been observed so far in tests of the equivalence principle or in the PPN formalism~(see \cite{will:1993fk,*will:2014la,uzan:2011vn} and references therein for a review of some tests of GR).

\section{Model}

In the following, we consider a general gravitational part of the action, such that the action reads
\begin{eqnarray}
	S&=&\frac{1}{c}\int d^4x \frac{\sqrt{-g}}{2\kappa}\left[f(\varphi)R-\frac{\omega(\varphi)}{\varphi}\left(\partial\varphi\right)^2\right] \nonumber\\
	&& \qquad + S_\textrm{mat} [g_{\mu\nu},\varphi,\Psi_i]  \, , \label{eq:actionstringframe}
\end{eqnarray}
with $\kappa=8\pi G/c^4$ and $S_\textrm{mat}=1/c \int d^4x\sqrt{-g}(\mathcal L_\textrm{SM}+\mathcal L_\textrm{int})$. The gravitational part of this action is slightly more general than the one used by Damour and Donoghue \cite{damour:2010zr,*damour:2010ve} that is characterized by $f(\varphi)=1$ and $\omega(\varphi)=2\varphi$. Except for the scalar-matter coupling, it corresponds to a generalized Brans-Dicke action \cite{brans:1961fk,*brans:2014sc}.

Damour and Donoghue have shown that the action used to model matter at the microscopic level (including the dilaton interaction) from Eq.~(\ref{eq:interLagrangeNL}) can phenomenologically be replaced at the macroscopic level by a standard point mass (or dust) action
\begin{equation}\label{eq:smat}
		S_\textrm{mat}[g_{\mu\nu},\varphi,\Psi_i]=-c^2\sum_A \int_A d\tau~ m_A(\varphi) \, ,
\end{equation}
where $d\tau$ is the proper time defined by $c^2d\tau^2=-g_{\alpha\beta}dx^\alpha dx^\beta$. Each mass $A$ can have its own composition such that the function $m_A(\varphi)$ will be different. The effects produced by the coupling of the dilaton to matter is encoded in the coupling function
\begin{equation}\label{eq:alpha}
	\alpha_A(\varphi)=\frac{\partial \ln m_A(\varphi)}{\partial \varphi}\, .
\end{equation}
Damour and Donoghue have derived a semi-analytical expression for the coupling $\alpha_A(\varphi)$. An approximation that is sufficient for our purpose writes \cite{damour:2010zr,*damour:2010ve}
\begin{equation}
	\alpha_A(\varphi)={D_g^*}'(\varphi) + \bar\alpha_A(\varphi)\, , \label{eq:alphasplit}
\end{equation}
where the prime denotes a derivative with respect to the scalar field and 
\begin{equation}\label{eq:DG*}
{D_g^*}(\varphi)=D_g(\varphi)+0.093\big(D_{\hat{m}}(\varphi)-D_g(\varphi)\big)+0.000 \ 27 D_e(\varphi) \, ,
\end{equation}
with $D_{\hat{m}}(\varphi)=(m_d D_{m_d}(\varphi)+m_u D_{m_u}(\varphi))/(m_d+m_u)$ and 
\begin{equation}\label{eq:bar_alpha_A}
\bar\alpha_A(\varphi)= \Big[\big(D_{\hat{m}}'(\varphi)-D_g'(\varphi)\big)Q'_{\hat m}+D_e'(\varphi)Q'_e\Big]_A\, .
\end{equation}
The coupling function $\alpha_A$ is explicitly written such that it is split into one composition independent term (${D_g^*}'$) and a term that is composition dependent ($ \bar\alpha_A$). Indeed, the dilatonic charges $[Q'_{\hat m}]_A$ and $[Q'_e]_A$ depends explicitly on the composition of the body $A$. Their expressions can be found in \cite{damour:2010zr,*damour:2010ve} (see notably table I from~\cite{damour:2010zr,*damour:2010ve}): $-Q'_{\hat m}$ ranges typically from $5\times 10^{-3}$ to $2\times 10^{-2}$ while $Q'_e$ ranges from $3\times 10^{-4}$ to $4\times 10^{-3}$. It has to be noted that the specific splitting in Eq. (\ref{eq:alphasplit}) between composition dependent and independent terms is no longer valid when the ratio of the atomic numbers $Z/A$ of the considered elements is not $1/2$. For simplicity, in this paper we shall assume that it is the case. The exact formula for Eq. (\ref{eq:alphasplit}) can be found in \cite{damour:2010zr,*damour:2010ve} from which other situations (e.g. nucleuses alone or light isotopes) can be worked out likewise.


\section{Einstein frame}

One can use a conformal transformation in order to write the action in a form where the kinetic part related to the metric is the one of GR. Such a conformal frame is called the Einstein frame. Moreover, one can rescale the scalar-field such that its kinetic part becomes canonical. In this frame, the action with the rescaled scalar-field reads
\begin{eqnarray} \label{eq:actionEinsteinframe}
 S&=&\frac{1}{c}\int d^4x \frac{\sqrt{-g_*}}{2\kappa_*}\left[R_* - 2\left(\partial_* \phi\right)^2\right] \nonumber \\
 &+& \frac{1}{c}\int d^4x \sqrt{-g_*}\mathcal L^*_\textrm{mat}\left[\frac{f_0}{f(\varphi)}g^*_{\mu\nu},\phi,\Psi_i \right]
\end{eqnarray}
where the stars denote Einstein frame quantities. The conformal transformation is defined by 
\begin{subequations}
 \begin{align}
    g_{\mu\nu}^* &= \frac{f(\varphi)}{f_0} g_{\mu\nu}, \qquad
    \mathcal L^*_\textrm{mat}= \left(\frac{f_0}{f(\varphi)}\right)^2 \mathcal L_\textrm{mat} \, ,\\
    \kappa_* &= \frac{\kappa}{f_0}  \, , \qquad \textrm{and	} \qquad \frac{d\phi}{d\varphi}=\sqrt{\frac{Z(\varphi)}{2}} \, ,
    \end{align}
where all quantities with a subscript 0 are evaluated at the background value of the dilaton, e.g. $f_0=f(\varphi_0)$ and	where the function $Z(\varphi)$ is defined by
	\begin{equation}
		Z(\varphi)=\frac{\omega(\varphi)}{\varphi f(\varphi)}+\frac{3}{2}\left(\frac{f'(\varphi)}{f(\varphi)}\right)^2 \, .\label{eq:Z}
	\end{equation}
\end{subequations}
The resulting field equations write \cite{damour:1992ys,damour:1994fk,damour:1994uq,hees:2016ax}
\begin{subequations}\label{eq:field_eq}
\begin{eqnarray}
  R^*_{\mu\nu} &=&\kappa_* \left(T_{\mu\nu}^*-\frac{1}{2}g^*_{\mu\nu} T^*\right) + 2 \partial_\nu \phi \partial_\mu\phi \, , \label{eq:metriceqEinstein} \\
  \Box_* \phi &=& -\frac{\kappa_*}{2}\sigma^* \, \label{eq:dilatonEF}
 \end{eqnarray}
\end{subequations} 
 where
 \begin{subequations}\label{eq:stress_energy}
 \begin{align}
 T^*_{\mu\nu}&=-\frac{2}{\sqrt{-g_*}}\frac{\delta \sqrt{-g_*}\mathcal L^*_\textrm{mat}}{\delta g^{\mu\nu}_*}=\sum_A \rho^*_A u^\mu_{*A} u^\nu_{*A}\\
\sigma^*&=\frac{\delta \mathcal L^*_\textrm{mat}}{\delta \phi}=-\sum_A \rho^*_A \alpha^*_A\, , \label{eq:sigmaEF}
\end{align}
with
\begin{equation}
	\rho^*_A=m^*_A\delta^{(3)}(\bm x-\bm x_A)/\sqrt{-g^*u^0_*}
\end{equation} 
the Einstein frame matter density for the body $A$ with the Einstein frame mass given by 
\begin{equation}
	m_A^*(\varphi)=\sqrt{f_0/f(\varphi)}m_A(\varphi)\, .
\end{equation}
\end{subequations}

The coupling appearing in the source term $\sigma^*$ (\ref{eq:sigmaEF}) of the Klein-Gordon equation (\ref{eq:dilatonEF}) is given by
\begin{eqnarray}
 \alpha^*_A(\phi)&=&\sqrt{\frac{2}{Z(\varphi)}}\left[ {D_g^*}'(\varphi)-\frac{1}{2}\frac{f'(\varphi)}{f(\varphi)}+\bar \alpha_A(\varphi)\right]\, , \ \  \label{eq:alphastar_phi} 
\end{eqnarray}

As one can see from Eqs. (\ref{eq:dilatonEF}) and (\ref{eq:sigmaEF}), $\alpha^*_A$ controls the coupling between matter and the dilaton in the field equations. Therefore, a decoupling appears when $\alpha^*_A$ is (or becomes) close to zero. For instance, Damour and Nordtvedt --- and later on Damour and Polyakov --- described a decoupling mechanism such that $\alpha^*_A$ is driven toward zero during the evolution of the universe \cite{damour:1993uq,*damour:1993kx,damour:1994fk,*damour:1994uq,*damour:2002ys,*damour:2002vn,barrow:1993pd,*barrow:1994pr,*serna:1996pr,*barrow:1997pr,*comer:1997pr,*mimoso:1998pl,*santiago:1998pr,*serna:2002ys,*jarv:2010uq,*jarv:2010fk,*jarv:2012vn,minazzoli:2014ao,*minazzoli:2014pb}. The mechanism is such that $Z(\varphi)$ is dynamically driven toward infinity as the universe expands. (Note that a qualitative description of such a mechanism had already been proposed by Steinhardt and Acceta in \cite{steinhardt:1990pl}). Here, we explore a different non-dynamical decoupling scenario in which the value of $\alpha_{A}^*$ is strongly reduced. Essentially,  decoupling scenarios arise when the first two terms in the square brackets of Eq. (\ref{eq:alphastar_phi}) cancel out (or almost do). As we shall see, this cancellation turns out to strongly reduce deviations from GR. With an exact cancellation, the whole phenomenology becomes similar to the GR one when the last term of Eq. (\ref{eq:alphastar_phi}) also tends toward 0.

\section{Post-Newtonian phenomenology}
Dilaton theories predict deviations for nearly all weak field observations at the Newtonian and post-Newtonian level. For example, the Newtonian equations of motion of a test mass orbiting a central mass (see \cite{hees:2016ax} for a detailed derivation) are given by
\begin{equation}\label{eq:eqnMotionN}
	\frac{d^2 \bm x_T}{dt^2}=-\frac{\tilde GM_A}{r_{AT}^3}\bm x_{AT}(1+\delta_T+\delta_{AT})\, ,
\end{equation}
where $\bm x_T$ is the vector position of the test mass $T$, $\bm x_{AT}=\bm x_T-\bm x_A$, $r_{AT}=|\bm x_{AT}|$ with $\bm x_A$ the vector position of the central body, $M_A$ is its mass and $\tilde G$ is the observed Newton constant which differs from the ones appearing in the action. The coefficients $\delta_T$ and $\delta_{AT}$ parametrize violations of the UFF. Their expressions are given by
\begin{equation}
	\delta_{T}=\frac{\alpha_0\tilde \alpha_{T0}}{1+\alpha_0^2}\, ,  \qquad 	\delta_{AT}=\frac{\tilde\alpha_{A0}\tilde \alpha_{T0}}{1+\alpha_0^2} \, ,
\end{equation}
where $\alpha_0$ is a universal constant 
\begin{equation}\label{eq:alpha0}
	\alpha_0=\sqrt{\frac{2}{Z_0}}\left[d_{g}^*-\frac{f'_0}{2f_0}\right]\, , \quad \textrm{with} \quad d_g^* = {D^*_g}'(\varphi_0) \, .\footnote{In what follows, we generically use the following abbreviation $d_X= D'_X(\varphi_0)$.}
\end{equation}
The coefficients $\tilde \alpha_{T0}$ depend explicitly on the composition of the body $T$
\begin{equation}
	\tilde \alpha_{T0}=\sqrt{\frac{2}{Z_0}}\bar\alpha_{T0} \, ,
\end{equation}
with $\bar  \alpha_T$ defined by Eq.~(\ref{eq:bar_alpha_A}). As we shall see, our decoupling scenarios are such that the two terms in the square brackets of Eq. (\ref{eq:alpha0}) almost cancel out such that one gets $\alpha_0 \sim 0$.

Let us note that, as far as we know, the parameters $\delta_{AB}$ have not been considered in the literature so far. In most cases, they can safely be neglected. Indeed, since the dilaton charges $-Q'_{\hat m}$ and $Q'_e$ have values lower than $10^{-2}$, the UFF violating parameters $\delta_{AT}$ are in general two orders of magnitude smaller than $\delta_T$. However, as we shall see, the parameters $\delta_{AT}$ become preponderant in some of the decoupling limits we are interested in. In those situations, the important point to notice is that they lead to a drastic modification of the UFF phenomenology. Indeed, while the parameters $\delta_T$ induce an additional term in the equation of motion (\ref{eq:eqnMotionN}) that depends on the composition of the test mass falling in the gravitational field, at the Newtonian level, the parameters $\delta_{AT}$ induce additional terms that depend on both the composition of the falling body and the composition of the source. As far as we know, this unusual prospect has not been considered neither from a theoretical nor from a phenomenological perspective so far. Therefore, this new type of phenomenology remain to be explored and constrained by observations.

The dilaton-matter coupling modifies also the behavior of atomic clocks. Observables relying on atomic clocks depends explicitly on the following parameters
\begin{equation}
	\hat\delta_I = \frac{\alpha_0\tilde \chi_{I0}}{1+\alpha_0^2}\, ,  \qquad 	\hat \delta_{IA}=\frac{\tilde\alpha_{A0}\tilde \chi_{I0}}{1+\alpha_0^2} \, ,
\end{equation}
where the quantity $\tilde \chi_{I0}$ depends on the atomic species and particular atomic transition used in the clock. As for the UFF violating term $\delta_{AT}$, $\hat \delta_{IA}$ becomes numerically significant in some decoupling limits only (see below). For clocks working on a hyperfine transition, and neglecting strange quark contributions, the $\tilde \chi_{I0}$ coefficient is given by \cite{hees:2016ax}
\begin{align}
		\tilde \chi_{I0} &=  \sqrt{\frac{2}{Z_0}}\Bigg[ -2 d_{m_e}+5.5\times10^{-4} (d_{m_e}-d_g)  \nonumber \\
	&-(4+K_{\textrm{rel}}-4.1\times10^{-4}) d_{e}- 0.0017 (d_{\delta m}-d_g) \nonumber\\
	&-  (\kappa_q+0.056) (d_{\hat m}-d_{g})\Bigg] \, ,\label{eq:S_tilde_chi} 
\end{align}
where $d_i=D'_i(\varphi_0)$, $d_{\delta m}=(m_d d_{m_d}-m_u d_{m_u})/(m_d-m_u)$ and $K_{\textrm{rel}}$ comes from the Casimir factor and reads (at the $s$-wave approximation\footnote{Numerical many-body calculations give more accurate results \cite{dzuba:1999pz}.}) $K_{\textrm{rel}}= (Z \af)^2 (12 \lambda^2-1)/(\lambda^2 (4 \lambda^2-1))^{-1}$, where $\lambda=[1-(\af  Z)^2]^{1/2}$, while $\kappa_q$ comes from the nuclear magnetic moment and is computed in \cite{flambaum:2006pr}. 

For example, the gravitational redshift between two clocks located at two different altitudes around Earth is given by
\begin{align}
	&\left.\frac{\Delta \nu}{\nu}\right|_\textrm{grav}=\left.\frac{\nu_A-\nu_B}{\nu_B}\right|_\textrm{grav} \nonumber\\
	=&\frac{ \tilde G M_\oplus}{r_A} \left(1+\hat\delta_A +\hat \delta_{A\oplus}\right)
	-\frac{ \tilde GM_\oplus}{r_B} \left(1+\hat\delta_B +\hat \delta_{B\oplus}\right) \, ,
\end{align}
where the subscripts $A/B$ refer to the positions and composition of the two clocks.

Similarly,  the frequency ratio of two clocks located at the same place will evolve with time because of the variation of the gravitational potential (see for instance \cite{guena:2012ys} and references therein). Considering only the gravitational potential of the Sun, dilaton theories predicts the following \cite{damour:1997qv,*damour:1999fk,nordtvedt:2002uq,hees:2016ax}
\begin{equation}
	\delta \left(\ln \frac{\nu_I}{\nu_J}\right) = \left(\hat \delta_I-\hat\delta_J +\hat\delta_{I\odot}-\hat\delta_{J\odot}\right) \frac{\delta W_\odot}{c^2}\, .
\end{equation}

 At the post-Newtonian level, the usual PPN formalism can be used \cite{damour:2010zr,hees:2016ax}. In particular, the universal $\gamma$ PPN coefficient measured for example with the Shapiro time delay  writes
\begin{equation}
	\gamma -1 = -\frac{2\alpha_0^2}{1+\alpha_0^2} \, .
\end{equation}

\section{Decoupling scenarios}
Different decoupling scenarios can arise in the general theory parametrized by the action (\ref{eq:actionstringframe}). The idea is to identify  theories that intrinsically produce a small value for the coupling $\alpha_A^*(\varphi)$ from Eq.~(\ref{eq:alphastar_phi}). In the following scenarios, observational deviations from GR are strongly reduced:

\begin{enumerate}	\item  $e^{2D_g(\varphi)} \propto f(\varphi)$: in this case, one has $\alpha_0=\sqrt{2/Z_0}\left(0.093(d_{\hat{m}}-d_{g})+0.000 \ 27 d_e\right)$.  Assuming $d_{\hat{m}}$, $d_{\hat{m}}-d_g$ and $d_{g}$ are of same order of magnitude, the value of $\alpha_0$ is reduced by one order of magnitude with respect to the general case --- see Eqs. (\ref{eq:alpha0}) and (\ref{eq:DG*}). It means that the Einstein Equivalence Principle (EEP) violating parameters $\delta_T$ and $\hat \delta_T$ are also reduced by about one order of magnitude. However, they still remain dominant compared to the other EEP violating parameters $\delta_{AT}$ and $\hat \delta_{AT}$. On the other hand, the deviation to unity of the PPN parameter $\gamma$ is reduced by two orders of magnitude.
	
		\item  $e^{2D_g(\varphi)}\propto e^{2D_{m_i}(\varphi)}\propto f(\varphi)$: in this case, one has $\alpha_0=\sqrt{2/Z_0}\, 0.000 \ 27 d_{e}$ and only the electromagnetic contribution plays a role in EEP violations. Assuming $d_{e}$ and $d_{g}$ are of same order  of magnitude, the EEP violating parameters $\delta_T$ and $\hat \delta_T$ are reduced by almost 4 orders of magnitude, while the PPN parameter $\gamma$ is reduced by around 7 orders of magnitude. Since the electromagnetic dilatonic charge $Q_e'$ is of order of $10^{-3}$ for most materials (see table I in \cite{damour:2010zr,*damour:2010ve}), the EEP violating parameters $\delta_{AT}$ and $\hat \delta_{AT}$ are no longer negligible compared to the EEP violating parameters $\delta_T$ and $\hat \delta_T$. Even more, they become predominant.
		
For instance, for test bodies made of platinum and titanium orbiting a celestial body made of silica (which is approximately the configuration of the MICROSCOPE mission \cite{touboul:2001kx,*touboul:2012cr,*berge:2015jp}) the UFF violating coefficients read $\delta_{Pt} \approx 2 \times 10^{-6} d_e^2/Z_0$, $\delta_{Pt ~SiO_2} \approx  10^{-5} d_e^2/Z_0$ and $\delta_{Ti} \approx  10^{-6} d_e^2/Z_0$, $\delta_{Ti~SiO_2} \approx  5 \times10^{-6}d_e^2/Z_0$. Therefore, the differential acceleration $\eta=\Delta a/a$ between the two test masses is given by~\citep{hees:2016ax} $\eta=6\times 10^{-6}d_e^2/Z_0$. This coefficient will be constrained at the level of $10^{-15}$ with MICROSCOPE~\cite{touboul:2001kx,*touboul:2012cr,*berge:2015jp}.

Similarly, considering a gravitational redshift test using a Caesium clock orbiting a celestial body made of silica (which is approximately the configuration of the ACES mission \cite{cacciapuoti:2011ve}), the EEP violating coefficients read $\hat \delta_{Cs} \approx -10^{-3} d_e(d_{m_e}+2.4 d_e)/Z_0$ and $\hat \delta_{Cs~SiO_2} \approx -5.4 \times 10^{-3} d_e(d_{m_e}+2.4 d_e)/Z_0$. The ACES mission will constrain the sum of these two parameters $-6.4 \times 10^{-3} d_e(d_{m_e}+2.4 d_e)/Z_0 $ at the level of $10^{-6}$~\cite{cacciapuoti:2011ve}. This shows the complementarity of both types of observations,  UFF tests being more sensitive to $d_e$ while gravitational redshift tests bring information on $d_e d_{m_e}$.
		
		\item $e^{2D_g(\varphi)}\propto e^{2D_{m_i}(\varphi)}$ and $d_e=0$:  in this case, all UFF violating parameters are vanishing but the PPN parameters $\gamma$ and $\beta$ are not equal to one in general (i.e. $\bar\alpha_{A0}=0$ but $\alpha^*_{A0}=\alpha_0 \neq 0$).
		
		\item $e^{2D_g(\varphi)}\propto e^{2D_{m_i}(\varphi)}\propto f(\varphi)$, and $d_e=0$: in this case the decoupling is total (i.e. $\alpha^*_{A0}=\alpha_0=\bar\alpha_{A0}=0$). All EEP violating parameters are vanishing and the PPN parameters $\gamma$ and $\beta$ are exactly equal to one --- regardless of the (non-singular) dilaton kinetic term in the original action (\ref{eq:actionstringframe}).
\end{enumerate}

Of course, there is a whole spectrum of decouplings in between the four simplified situations described above. Moreover, it is important to notice that such non-dynamical decouplings can come on top of the Damour and Nordtvedt dynamical decoupling \cite{damour:1993uq,*damour:1993kx,damour:1994fk,*damour:1994uq,*damour:2002ys,*damour:2002vn,barrow:1993pd,*barrow:1994pr,*serna:1996pr,*barrow:1997pr,*comer:1997pr,*mimoso:1998pl,*santiago:1998pr,*serna:2002ys,*jarv:2010uq,*jarv:2010fk,*jarv:2012vn,minazzoli:2014ao,*minazzoli:2014pb}. 

Among those decouplings, an intriguing case is the total decoupling scenario (fourth case). Indeed, it means that there can exist dilaton models with non-trivial non-minimal dilaton-matter coupling in the action that produce exactly the phenomenology of GR. Let us remind that in this situation, the dilaton-matter coupling term in the action reads 
\begin{eqnarray}
\label{eq:interLagrangeNLdecoupled}
\mathcal L_\textrm{int}=	D_g(\varphi)   &\Bigg[&-\frac{\beta_3}{2g_3}\left(F^A\right)^2\\
&&-\sum_{i=e,u,d}(1+\gamma_{m_i}) m_i\bar\psi_i\psi_i\Bigg]  . \nonumber
\end{eqnarray}
A close look at the equations in this case allows one to see that such models reduce at the macroscopic level to
\begin{equation} \label{eq:mafunctionfordecoupl}
 f(\varphi) \propto m_A^2(\varphi).
 \end{equation}
  Therefore, one recovers the decoupling studied in \cite{minazzoli:2013fk,*minazzoli:2015ax},  where a phenomenological multiplicative coupling between the scalar field and the rest mass energy density was assumed. The present study shows one specific way to justify the macroscopic phenomenology studied in \cite{minazzoli:2013fk,*minazzoli:2015ax} by a microphysics action of matter (\ref{eq:interLagrangeNL}). But it has to be noted that Eq. (\ref{eq:mafunctionfordecoupl}) equivalently reads as follows
 \begin{equation} \label{eq:planckmass}
M_{\textrm{Planck}}(\varphi) \propto m_A(\varphi) ,
 \end{equation}
where $M_{\textrm{Planck}}$ is the Planck mass. Therefore, one can explain the total decoupling by the fact that all dilatonic effects can be re-absorbed via a simple change of units. Therefore the theory is indeed indistinguishable from general relativity in the dust limit. In some sense, it is nothing but a ``Veiled'' general relativity \cite{deruelle:2011nd}. 

Although the decouplings presented in this paper seem to require some fine tuning of the dilaton-matter coupling (see e.g. Eq. (\ref{eq:planckmass})), one can conjecture that such apparent fine tunings may potentially arise from an underlying symmetry of nature. 

Let us note otherwise that with the effective relation Eq. (\ref{eq:mafunctionfordecoupl}), the whole action (\ref{eq:actionstringframe}) can be made conformally invariant for specific kinetic terms --- namely $f(\varphi)=\varphi^2$ and $\omega(\varphi)=-6~ \varphi$. However, it has to be stressed that the field equations are singular when the gravitational part of the action is invariant under conformal transformation. This can be witnessed through the vanishing value of the function $Z(\varphi)$ defined in (\ref{eq:Z}) in this case. This is in accordance with the fact that conformal invariance in scalar-tensor theories is a ``sham'' symmetry that does not add any degree of freedom compared to GR \cite{tsamis:1986ap,*jackiw:2015pr}. In other words, the decoupling studied in the present paper is not related to some conformal invariance of the gravitational part of the action density. 

\section{Conclusion and discussion}

In this paper, we showed that there exists a whole spectrum of non-dynamical decouplings between the dilaton and matter that arise directly from the non-minimal couplings in the action density. We studied the implications of such decouplings on the dilaton post-Newtonian phenomenology. Accordingly, one of the main finding of this article is to show that the equations of motion of a test mass orbiting a central mass have to be modified at the Newtonian level according to
\begin{eqnarray}
&&\frac{d^2 \bm x_T}{dt^2}=-\frac{\tilde GM_A}{r_{AT}^3}\bm x_{AT}(1+\delta_T) \\
&&\qquad \qquad \rightarrow \frac{d^2 \bm x_T}{dt^2}=-\frac{\tilde GM_A}{r_{AT}^3}\bm x_{AT}(1+\delta_T+\delta_{AT}). \nonumber
\end{eqnarray}
Indeed, although the parameters $\delta_{AT}$ are negligible with respect to the parameters $\delta_T$ in most models, they become preponderant in models that lead to some of the non-dynamical decouplings studied in this paper. As a consequence, such terms should be added when constraining the phenomenology of UFF violations from observations.

Because the decouplings are not dynamical, but rather take their roots from the specific form of the dilaton-matter and dilaton-Ricci couplings in the action density, they work whether the dilaton has a potential or not. In particular, they can happen for dilatonic dark matter models \cite{damour:1990pl,stadnik:2015pd,*stadnik:2015pu,*Stadnik:2016ab,arvanitaki:2015pr,*arvanitaki:2016pl,*arvanitaki:2016ar,hees:2016ar} as well. However, a self-interaction of the dilaton can lead to departure (with various strength) from general relativity's phenomenology, even if the dilaton is not (or weakly) sourced by matter fields \cite{barrow:1993pd,*barrow:1994pr,*serna:1996pr,*barrow:1997pr,*comer:1997pr,*mimoso:1998pl,*santiago:1998pr,*serna:2002ys,*jarv:2010uq,*jarv:2010fk,*jarv:2012vn,minazzoli:2014ao,*minazzoli:2014pb}.

It has to be noted that the decoupling scenarios require various degrees of apparent fine-tuning in the action density. The most obvious example is for the total decoupling scenario that turns out to originate from the simple relation between the Planck mass and all the other particles' mass (see Eq. (\ref{eq:planckmass})). The only way to avoid this shortcoming is to conjecture that one of the apparent tunings hypothetically originates from something more fundamental. For instance, the theory presented in \cite{ludwig:2015pl} may be one possible justification from a fundamental perspective. There, the universal coupling would come from the way matter and geometry are made intrinsically inseparable in an original $f(R,\mathcal{L}_m)$ action. Otherwise, let us note that a universal coupling is also one of the conditions that are required in order to have a dynamical decoupling during the evolution of the universe as well \cite{damour:1994fk,*damour:1994uq}.

It has to be stressed however that in this study we used the usual dust field approximation in order to model matter. The reason is that one knows how to make the link between the (semi-)fundamental matter Lagrangian from Eq.~(\ref{eq:interLagrangeNL}) and its effective point particles realization from Eq.~(\ref{eq:smat}), thanks notably to the works of Damour and Donoghue \cite{damour:2008pr,damour:2010zr,*damour:2010ve}. Nevertheless, the non-minimal scalar-matter coupling can eventually lead to non-trivial effects when considering pressureful fluids. For instance, it has been argued that taking into account the pressure can soften the various decouplings in comparison to the dust case considered in this paper \cite{minazzoli:2013fk,*minazzoli:2015ax,minazzoli:2014ao,*minazzoli:2014pb}. Extending the results of the present study to pressureful cases could be an important step. However several theoretical issues have to be addressed first. Indeed, because of the non-minimal scalar-matter coupling in the action density, parts of the material Lagrangian appear explicitly in the field equations and it is a difficult task to derive from first principles the value of these on-shell parts when one eventually wants to consider effective pressureful fluids. This problem is left for further studies.

Also, taking into account quantum electrodynamics (QED) effects, through the QED trace anomaly, may shift the space of parameters \cite{nitti2012pr}, such that the decoupling would be total for $e^{2D_g(\varphi)}\propto e^{2D_{m_i}(\varphi)}\propto e^{2D_{e}(\varphi)}\propto f(\varphi)$, instead of $e^{2D_g(\varphi)}\propto e^{2D_{m_i}(\varphi)}\propto f(\varphi)$, and $d_e=0$. However, the splitting between purely QCD and QED effects is ambiguous when QED is turned on \cite{gasser:2003ep}. Therefore a precise derivation of the QED and QCD contributions to the particles' mass demands involved quantum field theory calculations that are out of the scope of this paper. 

Finally, it has been shown in \cite{armendariz:2012pr}, that once the fields with gravitational strength are included in the quantum loops, the equivalence principle is violated at the quantum level, even if the scalar field $\varphi$ does not appear in the material Lagrangian (i.e. $D_X(\varphi)=D_X$ in (\ref{eq:interLagrangeNL})). The reason behind this is that no symmetry protects the structure of the scalar-matter coupling of the original action, such that radiative corrections generate terms with explicit couplings between the scalar field $\varphi$ and matter, while the coupling between the metric and matter is protected by diffeomorphism invariance \cite{armendariz:2012pr}. However, those corrections are expected to be extremely small. For instance, when $D_X(\varphi)=D_X$, they have been estimated to be at best proportional to $(\kappa)^{-3/2}\sim M_P^{-3}$, where $M_P$ is the Planck mass \cite{armendariz:2012pr}.

\begin{acknowledgments}
The authors are thankful to Peter Wolf and Quentin Bailey for their helpful comments on a preliminary version of this manuscript. The authors are also grateful to John Donoghue, Federico Piazza, Francesco Nitti, Cristian Armendariz-Picon, Gilles Esposito-Far\`ese and Martin Hoferichter for their helpful discussions. 

\end{acknowledgments}

%

\end{document}